\documentclass{article}
\usepackage{spconf,amsmath,graphicx}
\usepackage{xcolor}
\usepackage{floatrow}

\title{Comparing Deep Learning strategies for paired but unregistered multimodal segmentation of the liver in T1 and T2-weighted MRI}
 \name{Vincent Couteaux$^{\star \dagger}$ \qquad Mathilde Trintignac$^{\ddag}$ \qquad Olivier Nempont $^{\dagger}$
 \qquad Guillaume Pizaine $^{\dagger}$} 
 \names{Anna Sesilia Vlachomitrou $^{\dagger}$
 \qquad Pierre-Jean Valette $^{\ddag}$\qquad Laurent Milot $^{\ddag}$\qquad Isabelle Bloch $^{\star}$}

 \address{$^{\star}$LTCI, T\'el\'ecom Paris, Institut Polytechnique de Paris, France \\ $^{\dagger}$Philips Research Paris, Suresnes, France\\
Hospices Civils de Lyon, Lyon, France}

\newcommand{\VC}[1]{} 

\begin{document}
\ninept

\maketitle
\begin{abstract}
We address the problem of multimodal liver segmentation in paired but unregistered T1 and T2-weighted MR images. 
We compare several strategies described in the literature, with or without multi-task training, with or without pre-registration. 
We also compare different loss functions (cross-entropy, Dice loss, and three adversarial losses).
All methods achieved comparable performances with the exception of a multi-task setting that performs both segmentations at once, which performed
poorly.

\end{abstract}
\section{Introduction}
\label{sec:intro}

Automatic liver segmentation tools are used in a clinical context to ease the interpretation of medical images, 
as well as for quantitative measurements.
For instance, they are used for volumetry or as a preliminary step for automated assessment of hepatic fat fraction or liver tumor burden. 
Such analyses can be done with multiple modalities, among which T1 and T2-weighted MRIs\VC{, which give useful and complementary information for the diagnosis of hepatic lesions}.

\begin{figure}[t]
    \centering
    \includegraphics[width=0.98\textwidth]{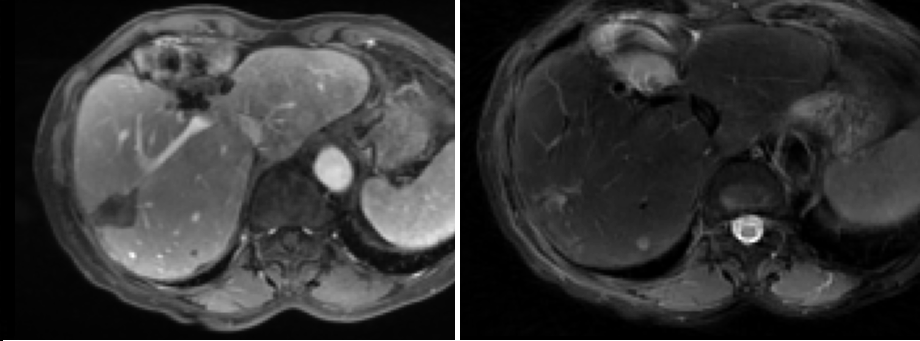}
    \caption{Multisequence MRI pair example. Left: T1-weighted image. Right: T2-weighted image.}
    \label{fig:images}
\end{figure}

T1 weighted MRIs are anatomical images, and thus show a well-contrasted and easy to segment liver.
Conversely, T2-weighted images have a lower inter-slice resolution. They are more prone to acquisition artifacts, and the liver is harder to distinguish from its surroundings (see Figure~\ref{fig:images}).
Therefore, the accurate manual segmentation of the liver is more tedious in T2 images.
T1 and T2-weighted images are acquired a few minutes apart, which induces misalignments
between the two images (mainly due to breathing). 

In this work, \VC{rather than introducing a novel method}, we propose a comparison of different strategies described in the literature to perform accurate automatic segmentation of the liver in pairs of T1 and T2-weighted MRI images.


In the recent literature, most proposed segmentation methods are built upon the U-Net architecture~\cite{ronneberger2015u}, 
either by
integrating shape priors \cite{Zeng2019liver},
modifying the architecture \cite{Li2018HDenseUNetHD},
or adapting to a particular problem,
such as joint lesion segmentation \cite{Vorontsov2018LiverLS}, unsupervised domain adaptation \cite{Yang2019UnsupervisedDA}, or integration of manual corrections \cite{Chlebus2019ReducingIV}.
We therefore chose this setting as a basis for our comparison.


\begin{figure*}[t]
    \centering
    \includegraphics[width=0.9\textwidth]{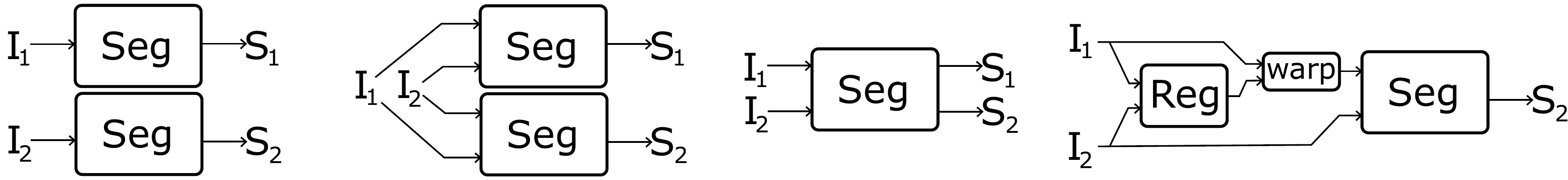}
    \caption{Inference of a pair of images for different input/output strategies. Left to right: single input; double-input, single-output; double-input, double-output; double-input, single-output preregistered. 
    For the {\em specialized} setting, $I_1$ is the T1 image of the pair and $I_2$ the T2 image. 
    For the {\em unspecialized} setting,
    the T1 and T2 images are randomly swapped during training, so that the networks do not expect a particular modality in each channel.
    }
    \label{fig:strategies}
\end{figure*}

 Our experiments revolve around two axes.
The first axis is the multi-input and multi-output strategy:
it is shown in~\cite{Wang2019AutomatedCA} that it is possible to train efficient
unspecialized segmentation networks when no paired data are available,
and many works showed the usefulness of dual-input segmentation for registered paired images~\cite{Li2018HDenseUNetHD,Zhou2019ARD}.
We investigate to what extent these strategies are applicable to our particular problem and data set, and how they compare. The goal is to answer the following questions:
Does providing both images of the pair as input to the network 
increases performance?
If so, does segmenting both images in one forward pass, in a multi-task way, enables to get better results?
Can we improve the results by applying a non-linear registration algorithm to the images before forwarding them to the network?
Our hypothesis is that, by providing anatomical information through the T1 image, a segmentation network may succeed better in segmenting the T2 image, even with misalignements.

The second axis is the objective function we optimize during training. The importance of this parameter in segmentation has been demonstrated in~\cite{jadon2020survey,sudre2017generalised}, and the benefit of learned segmentation losses by adversarial training for certain applications in~\cite{Samson2019IBY,Luc2016SemanticSU,ghafoorian2018elgan}.
We compare three standard voxel-to-voxel loss functions, and three other functions taking advantage of adversarial training.
For this axis the hypothesis is that adversarial losses, by learning the distribution of liver masks, may lead to more robust predictions and an increased performance overall.


\section{Data}
\label{sec:data}

We compare the different approaches on a database of
88 pairs of T1-weighted and T2-weighted MRIs centered on the liver, coming from 51 patients which all have hepatic lesions. 
The T1 images are acquired at a portal enhancement time.
We resize every image at a resolution of 3 mm along the $z$ axis, and 1.5 mm along the $x$ and $y$ axes.
Images in each pair are aligned with their DICOM metadata, but breathing can induce large movements of the liver (up to $15$cm).

Reference segmentation masks are obtained through manual annotations by a radiologist,
using interactive 3D tools. Note that due to low contrast of the liver in T2 images, as well as lower resolution along the $z$ axis, manual annotation of the liver in T2 images is difficult and less accurate than in T1 images.
We split the dataset by keeping 12 pairs for testing, and 6 pairs for validation.

\section{Methods}
\label{sec:method}

\subsection{Architecture and training}
In order to focus on the strategies and the objective functions, and for a fair comparison, we let the following architecture and optimization parameters fixed.
We use the 3D-U-net architecture \cite{ronneberger2015u,milletari2016v}, pre-trained with weights provided in \cite{zhou2019models}, as this architecture is now standard
for medical image segmentation \cite{Isensee2018nnUNetSF}.
We use the Adam optimizer at a $1.10^{-4}$ learning rate with early-stopping,
by keeping the best network on the validation dataset, for 900 epochs of 150 steps.
For memory reasons, we use batches of size 1, and crop inputs into cuboids of random sizes and ratios. 
We use a random intensity shift as a data augmentation strategy.

\subsection{Multi-channel and multi-output}
\label{subsec:strategy}

We evaluate the benefit of multi-modal inputs by comparing different input/output settings (see Figure~\ref{fig:strategies}).
Each setting has two versions, which we refer to as {\em specialized} or {\em unspecialized} and that we describe below.

\begin{description}
    \item[Single input:]
    The network has only one channel for input and output.
    If {\em specialized}, two networks are trained, one for each modality. If {\em unspecialized}, only one network is trained, taking as input indifferently a T1 or T2 image.
    \item[Double input, single output:]
    The network has two channels as input, and one as output.
    We train it to segment the image in the first channel, the second channel receiving the auxiliary modality.
    If {\em specialized}, two networks are needed, one segmenting T1 images with T2 as auxiliary modality, and vice versa.
    If {\em unspecialized}, only one network is needed, segmenting indifferently T1 or T2 images.
    \item[Double input, double output:] 
    A single network is trained, where both the input and output layers have two channels - one for each image of the pair - so that both predictions are multi-tasked. When {\em specialized}, the first channel is always the T1 image and the second channel always T2, whereas we randomly swap the channels during traing when {\em unspecialized}, so that the network does not expect a particular modality in each channel.
    \item[Pre-registered:]
    To study how registering the images before the segmentation may be beneficial,
    we performed a non-linear registration of the T1 image on the T2 image of each pair and trained a double input, single output network segmenting T2 images with T1 as auxiliary modality.
    As the T2 image is harder to segment, we can expect that by also providing the aligned T1 image to the network, it can use all relevant information to make a more accurate segmentation.
    
\end{description}

\subsection{Influence of the loss fonction}

In the following section we consider a dataset of images $\{x_n\} \subset X$ and annotations $\{y_n\} \subset Y$, and a segmentor $S : X \rightarrow Y$.
To evaluate the influence of the loss function on the performances, we test 6 of them
on the single-input, unspecialized strategy.

Three loss functions are standard voxel-to-voxel functions: 
\begin{description}
    \item[Binary cross-entropy:]
        $L_{bce}(x, y) = -\sum y\log(S(x)) $, where the sum is taken over all image voxels, as proposed 
        in \cite{ronneberger2015u}.
    \item[Dice:]
        $L_{Dice}(x, y) = -2\sum yS(x)/(\sum y \sum S(x)) $, as proposed in~\cite{sudre2017generalised}. The normalization enables a better performance
        for important class imbalance between foreground and background ({\em i.e.} when the objects to segment are small) compared to $L_{bce}$.
    \item[Cross-entropy+Dice:]
        $L_{sum} = L_{bce} + L_{Dice}$
\end{description}

Three loss functions are adversarial losses: we simultaneously train a discriminator network $D : Y \rightarrow [0,1]$ to recognize reference masks.
The idea is that by learning the distribution of liver masks, we can get models that are robust to improbable predictions (with holes or odd shape for instance).
This technology has gained popularity in segmentation 
\cite{Hung2018AdversarialLF,Zhang2017DeepAN},
and we refer to the {\em related works} section in~\cite{Samson2019IBY} for a good review of adversarial learning in segmentation.

\begin{description}
    \item[The vanilla GAN loss,] as in \cite{Luc2016SemanticSU}:
        $ L_{GAN}(x,y) = L_{bce}(S(x),y)) + L_{bce}(D(S(x)), 1) $
    \item[The embedding loss,] proposed in \cite{ghafoorian2018elgan}:
        $L_{el}(x,y) = L_{bce}(S(x), x)   + || D_k(S(x)) - D_k(y) ||^2$, 
        where $D_k$ represents the $k$-th layer of network $D$.
        The goal is to gain in stability during training.
    \item[The gambling loss,] described in \cite{Samson2019IBY}:
    Instead of the discriminator $D$, we use a gambler $G : X \times Y \rightarrow Y $, which takes as input an image and a segmentation, and outputs a normalized betting map.
    It is trained to minimize
    $L_G(x,y) = - \sum G(x, S(x))y\log(S(x))$
    while the segmentor minimizes
    $L_S = L_{bce} - L_G$
    The goal is to train
    the gambler network able to recognize the hard part of the image to segment, so that the segmentor focuses on them.
    
\end{description}


We start the training with a pre-trained segmentor, with the single-input unspecialized strategy
(see Section~\ref{subsec:strategy}),
as we found that it performed as well as any other (see Section~\ref{sec:results}).

\section{Results, discussion and conclusions}
\label{sec:results}

We report the performance of all approaches using Dice coefficient.
Hausdorff distance and Hausdorff distance 95th percentile resulted in a similar ranking of the approaches.
To help with the interpretation of the results, we perform a 4-fold cross-validation on the entire database using the single-input unspecialized strategy, that we repeat 3 times.
On average, the Dice scores of the different runs on each fold differ by $0.005$.
Any Dice score difference below that will thus not be considered meaningful for comparing strategies.

Table \ref{tab:strat} shows a comparison of performances among the training strategy, recorded on the test dataset.
We note a slightly worse performance for T2 images in specialized strategies
compared to their unspecialized counterpart,
while the difference of Dice scores of T1 images is limited. 
Multi-output networks performed significantly worse than multi-input ones.
As shown in \cite{Wu2020Understanding}, multi-task learning is not trivial
to work with, and the naive approach we tried showed its limits;
it may be especially tricky in
3D segmentation, where increasing the capacity of a network is costly in terms of memory.

Let us now assess how much a network relies on the auxiliary modality for double-input settings.
To this end, we evaluate the performance of a network when segmenting an image with its matching auxiliary image, and compare it to its performance when given mismatched pairs.
We measure this performance gap with the difference of Dice scores, and compile the results in Table \ref{tab:gain}.
We record a more important gap for T2 images, which is consistent with the idea that, as this modality is harder to segment, the network learns to fetch information in the T1 modality.
This gap is even more important when images are pre-registered, which is also expected as the information is easier to retrieve when images are aligned.
Multi-output settings showed the most important need for the auxiliary modality.
Contrary to our hypothesis according to which adding T1 information should improve the T2 predictions,
comparing Table~\ref{tab:strat} with Table~\ref{tab:gain} seems to link an increased usage of the auxiliary channel with bad performances.
One explanation may be that the network may rely on irrelevant pieces
of additional information.
Overall, we found that T2 images contain enough information to accurately segment the liver, and the gap in performance between T1 and T2 images can be explained by a difference in annotation quality.

Table~\ref{tab:loss} shows the performance of networks trained with different loss functions.
We found no clear effect of this parameter on performance.
Adversarial training did not outperform the other loss functions either,
despite an expected behavior during the adversarial training and good discriminator performance. 
This experiment corroborates the idea expressed in \cite{Isensee2018nnUNetSF,hofmanninger2020automatic} that
data variety and annotation quality are prominent,
and that the gains induced by those innovations are sometimes hard to replicate on custom datasets.
Our cross-validation experiment showed high score differences across splits,
and showed that the split we chose for comparing strategies was particularly difficult (one cross-validation split averages a Dice of $0.983$ on both T1 and T2). This also corroborates the idea of prominence of data on performance.

\begin{figure*}[h]
    \centering
    \includegraphics[width=0.75\textwidth]{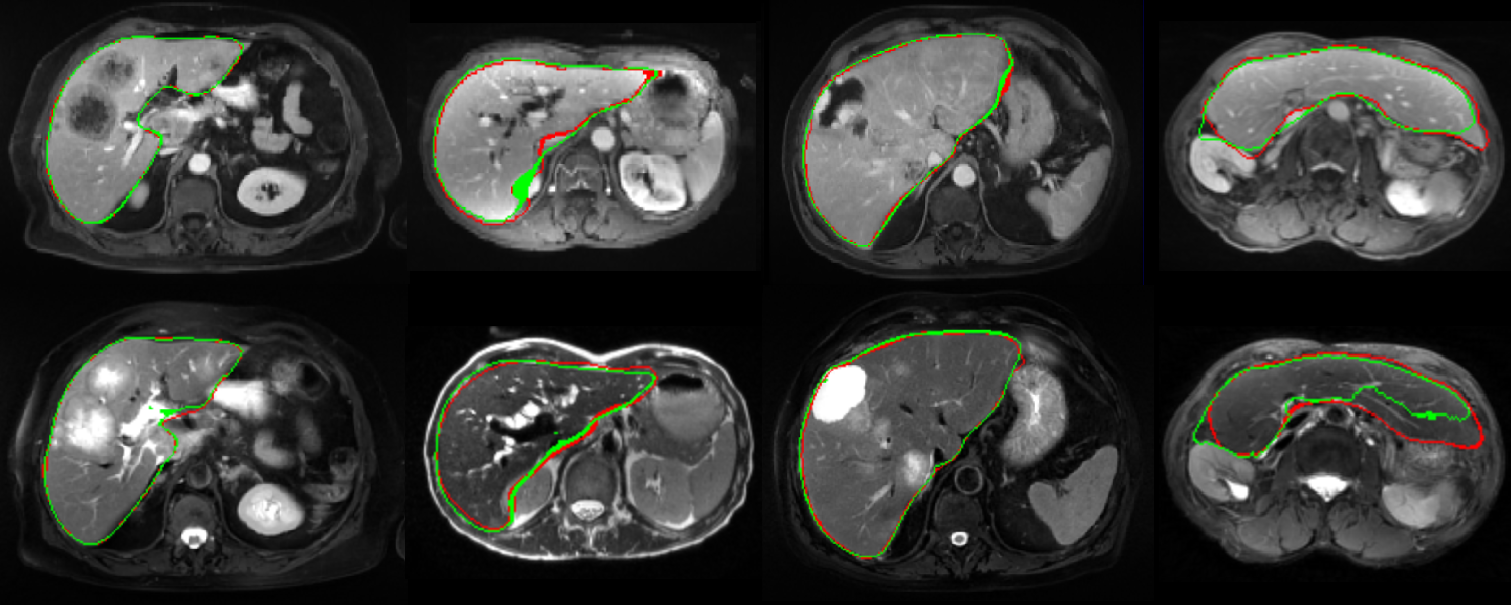}
    \caption{A few examples from the test database. Top: T1 image, bottom: T2 image.
    Red: manual annotation, green: prediction from the single-input, unspecialized network.}
    \label{fig:result}
\end{figure*}

Figure~\ref{fig:result} shows some examples of predictions of the single-input, unspecialized method. 
We can see that the predictions remain accurate even in the presence of large tumor burden (leftmost column) or big lesions near the edge of the liver (middle right column). 
The rightmost column shows a case were the patient has undergone an hepatectomy.
As it is the only case with such atypical anatomy on our database, it is particularly challenging.
Despite this difficulty the network managed to make accurate predictions, especially on the T1 image.


\begin{table}
    \centering
\ttabbox{
\begin{tabular}{l||c|c||c|c}
            & \multicolumn{2}{c}{Unspec.}
            & \multicolumn{2}{c}{Spec.}\\
     Strategy & T1 & T2 & T1 & T2 \\
     \hline
     1 input & 0.961 & 0.938 & 0.959 & 0.929 \\
     2 in, 1 out & 0.955 & 0.933 & 0.954 & 0.929 \\
     2 in, 2 out & 0.938 & 0.907 & 0.942 & 0.897 \\
     1 out, prereg. & - & - & - & 0.925
\end{tabular}
}{ \caption{Mean Dice scores vs. training strategy.}  \label{tab:strat}}

\ttabbox{
    \caption{Dice loss when pairs are mismatched vs. strategy.}
    \centering
\begin{tabular}{l||c|c||c|c}
            & \multicolumn{2}{c}{Unspec.}
            & \multicolumn{2}{c}{Spec.}\\
     Strategy & T1 & T2 & T1 & T2 \\
     \hline
     2 in, 1 out & $<$0.001 & 0.016 & 0.001 & 0.030\\
     2 in, 2 out & 0.042 & 0.083 & 0.023 & 0.103 \\
     1 out, prereg. & - & - & - & 0.054
\end{tabular}
  }{  \caption{Dice reduction when pairs are mismatched vs. strategy.} \label{tab:gain}}

\ttabbox{
\centering
\begin{tabular}{c||c|c}
      Loss & T1 & T2\\
      \hline
    Binary cross-entropy &  0.961 & 0.938 \\
    Dice loss & 0.959 & 0.931 \\
    Dice + BCE & 0.959 & 0.932 \\
    Vanilla GAN & 0.950 & 0.930 \\
    Embedding & 0.960 & 0.935 \\
    Gambling & 0.959 & 0930
\end{tabular}
}{\caption{Mean Dice scores vs. loss function.} \label{tab:loss}}
\end{table}

\bibliographystyle{IEEEbib}
\bibliography{strings,refs,biblio}

\begin{thebibliography}{10}

\bibitem{ronneberger2015u}
Olaf Ronneberger, Philipp Fischer, and Thomas Brox,
\newblock ``U-net: Convolutional networks for biomedical image segmentation,''
\newblock in {\em International Conference on Medical image Computing and
  Computer-Assisted Intervention}. Springer, 2015, pp. 234--241.

\bibitem{Zeng2019liver}
Qi~Zeng, Davood Karimi, Emily~HT Pang, Shahed Mohammed, Caitlin Schneider,
  Mohammad Honarvar, and Septimiu~E Salcudean,
\newblock ``Liver segmentation in magnetic resonance imaging via mean shape
  fitting with fully convolutional neural networks,''
\newblock in {\em International Conference on Medical Image Computing and
  Computer-Assisted Intervention}. Springer, 2019, pp. 246--254.

\bibitem{Li2018HDenseUNetHD}
Xiaomeng Li, Hao Chen, Xiaojuan Qi, Qi~Dou, Chi-Wing Fu, and Pheng-Ann Heng,
\newblock ``{H-DenseUNet}: hybrid densely connected {UNet} for liver and tumor
  segmentation from {CT} volumes,''
\newblock {\em IEEE Transactions on Medical Imaging}, vol. 37, no. 12, pp.
  2663--2674, 2018.

\bibitem{Vorontsov2018LiverLS}
Eugene Vorontsov, An~Tang, Chris Pal, and Samuel Kadoury,
\newblock ``Liver lesion segmentation informed by joint liver segmentation,''
\newblock in {\em IEEE 15th International Symposium on Biomedical Imaging
  (ISBI)}. IEEE, 2018, pp. 1332--1335.

\bibitem{Yang2019UnsupervisedDA}
Junlin Yang, Nicha~C Dvornek, Fan Zhang, Julius Chapiro, MingDe Lin, and
  James~S Duncan,
\newblock ``Unsupervised domain adaptation via disentangled representations:
  Application to cross-modality liver segmentation,''
\newblock in {\em International Conference on Medical Image Computing and
  Computer-Assisted Intervention}. Springer, 2019, pp. 255--263.

\bibitem{Chlebus2019ReducingIV}
Grzegorz Chlebus, Hans Meine, Smita Thoduka, Nasreddin Abolmaali, Bram van
  Ginneken, Horst~Karl Hahn, and Andrea Schenk,
\newblock ``Reducing inter-observer variability and interaction time of {MR}
  liver volumetry by combining automatic {CNN}-based liver segmentation and
  manual corrections,''
\newblock {\em PloS one}, vol. 14, no. 5, pp. e0217228, 2019.

\bibitem{Wang2019AutomatedCA}
Kang Wang, Adrija Mamidipalli, Tara Retson, Naeim Bahrami, Kyle Hasenstab,
  Kevin Blansit, Emily Bass, Timoteo Delgado, Guilherme Cunha, Michael~S
  Middleton, et~al.,
\newblock ``Automated {CT} and {MRI} liver segmentation and biometry using a
  generalized convolutional neural network,''
\newblock {\em Radiology: Artificial Intelligence}, vol. 1, no. 2, pp. 180022,
  2019.

\bibitem{Zhou2019ARD}
Tongxue Zhou, Su~Ruan, and St{\'e}phane Canu,
\newblock ``A review: Deep learning for medical image segmentation using
  multi-modality fusion,''
\newblock {\em Array}, vol. 3-4, pp. 100004, 2019.

\bibitem{jadon2020survey}
Shruti Jadon,
\newblock ``A survey of loss functions for semantic segmentation,''
\newblock {\em ArXiv:2006.14822}, 2020.

\bibitem{sudre2017generalised}
Carole~H Sudre, Wenqi Li, Tom Vercauteren, Sebastien Ourselin, and M~Jorge
  Cardoso,
\newblock ``Generalised dice overlap as a deep learning loss function for
  highly unbalanced segmentations,''
\newblock in {\em Deep learning in medical image analysis and multimodal
  learning for clinical decision support}, pp. 240--248. Springer, 2017.

\bibitem{Samson2019IBY}
Laurens Samson, Nanne van Noord, Olaf Booij, Michael Hofmann, Efstratios
  Gavves, and Mohsen Ghafoorian,
\newblock ``I bet you are wrong: Gambling adversarial networks for structured
  semantic segmentation,''
\newblock in {\em IEEE International Conference on Computer Vision Workshops},
  2019.

\bibitem{Luc2016SemanticSU}
Pauline Luc, Camille Couprie, Soumith Chintala, and Jakob Verbeek,
\newblock ``Semantic segmentation using adversarial networks,''
\newblock {\em ArXiv:1611.08408}, 2016.

\bibitem{ghafoorian2018elgan}
Mohsen Ghafoorian, Cedric Nugteren, N{\'o}ra Baka, Olaf Booij, and Michael
  Hofmann,
\newblock ``El-gan: Embedding loss driven generative adversarial networks for
  lane detection,''
\newblock in {\em European Conference on Computer Vision (ECCV)}. 2018, pp.
  256--272, Springer.

\bibitem{milletari2016v}
Fausto Milletari, Nassir Navab, and Seyed-Ahmad Ahmadi,
\newblock ``V-net: Fully convolutional neural networks for volumetric medical
  image segmentation,''
\newblock in {\em Fourth International Conference on 3D Vision (3DV)}. IEEE,
  2016, pp. 565--571.

\bibitem{zhou2019models}
Zongwei Zhou, Vatsal Sodha, Md~Mahfuzur~Rahman Siddiquee, Ruibin Feng, Nima
  Tajbakhsh, Michael~B Gotway, and Jianming Liang,
\newblock ``Models genesis: Generic autodidactic models for {3D} medical image
  analysis,''
\newblock in {\em International Conference on Medical Image Computing and
  Computer-Assisted Intervention}. Springer, 2019, pp. 384--393.

\bibitem{Isensee2018nnUNetSF}
Fabian Isensee, Jens Petersen, Andr{\'e} Klein, David Zimmerer, Paul~F. Jaeger,
  onnon Kohl, Jakob Wasserthal, Gregor Koehler, Tobias Norajitra, Sebastian~J.
  Wirkert, and Klaus Maier-Hein,
\newblock ``{nnU-Net}: Self-adapting framework for {U-Net}-based medical image
  segmentation,''
\newblock {\em ArXiv:1809.10486}, 2018.

\bibitem{Hung2018AdversarialLF}
Wei-Chih Hung, Yi-Hsuan Tsai, Yan-Ting Liou, Yen-Yu Lin, and Ming-Hsuan Yang,
\newblock ``Adversarial learning for semi-supervised semantic segmentation,''
\newblock in {\em BMVC}, 2018.

\bibitem{Zhang2017DeepAN}
Yizhe Zhang, Lin Yang, Jianxu Chen, Maridel Fredericksen, David~P Hughes, and
  Danny~Z Chen,
\newblock ``Deep adversarial networks for biomedical image segmentation
  utilizing unannotated images,''
\newblock in {\em International Conference on Medical Image Computing and
  Computer-Assisted Intervention}. Springer, 2017, pp. 408--416.

\bibitem{Wu2020Understanding}
Sen Wu, Hongyang~R. Zhang, and Christopher Ré,
\newblock ``Understanding and improving information transfer in multi-task
  learning,''
\newblock in {\em International Conference on Learning Representations}, 2020.

\bibitem{hofmanninger2020automatic}
Johannes Hofmanninger, Forian Prayer, Jeanny Pan, Sebastian R{\"o}hrich, Helmut
  Prosch, and Georg Langs,
\newblock ``Automatic lung segmentation in routine imaging is primarily a data
  diversity problem, not a methodology problem,''
\newblock {\em European Radiology Experimental}, vol. 4, no. 1, pp. 1--13,
  2020.

\end{thebibliography}

\section{Acknowledgments}

This work has been partially funded by a grant from {\em Association Nationale de la Recherche et de la Technologie} (\#2018/0199).

\section{Conflict of interest}

The authors have no relevant financial or non-financial interests to disclose.

\section{Compliance with Ethical Standards}



This study was performed in line with the principles of the Declaration of Helsinki. 
This is conform to standard reference methodology MR-004 of the CNIL (France).
Approval was granted by the CNIL (Study number 19-188).


\end{document}